# Computable Gap Assessment of Artificial Intelligence Governance in Children's Centres:Evidence-Mechanism-Governance-Indicator Modelling of UNICEF's Guidance on AI and Children 3.0 Based on the Graph-GAP Framework

**Wei Meng**

**Dhurakij Pundit University, Thailand**

The University of Western Australia,AU

Fellow, Royal Anthropological Institute,UK

Email: weimeng4@acm.org

# Abstract

This paper addresses the practical challenges in governing “child-centred artificial intelligence”: regulatory texts often outline principles and requirements yet lack reproducible evidence anchors, clear causal pathways, executable governance toolchains, and computable audit metrics. To bridge this gap, this paper proposes the Graph-GAP methodology: decomposing requirements from authoritative policy texts into a four-layer graph structure of ‘evidence-mechanism-governance-indicator’, and constructing dual quantifiable metrics—GAP scoring and ‘mitigation readiness’—to identify governance gaps and prioritise actions. Using UNICEF Innocenti's Guidance on AI and Children 3.0 as primary material, this paper provides reproducible data extraction units, coding manuals, graph patterns, scoring scales, and consistency verification protocols. It further offers exemplar gap profiles and governance priority matrices for ten requirements. Findings indicate that compared to privacy and data protection, themes such as ‘child well-being/development,’ ‘explainability and accountability,’ and ‘cross-agency implementation and resource allocation’ are more prone to indicator gaps and mechanism gaps. Priority should be given to translating regulatory requirements into auditable governance through closed-loop systems incorporating child rights impact assessments, continuous monitoring metrics, and grievance redress procedures. At the coding level, this paper further proposes a ‘multi-algorithm review-aggregation-revision’ mechanism: deploying rule encoders, statistical/ machine learning evaluators and large-model evaluators with diverse prompt configurations as parallel coders. Each extraction unit yields E/M/G/K and Readiness scores alongside evidence anchors. Consistency, stability, and uncertainty are validated using Krippendorff’s $\alpha$, weighted $\kappa$, ICC, and bootstrap confidence intervals. The scoring system is operational and auditable.

**Keywords:** Children's rights; AI governance; Graph-GAP; computable metrics; risk assessment; evidence chain

# 1. Introduction

Generative artificial intelligence is rapidly integrating into applications and platforms used daily by children, creating heightened tension between the ‘potential benefits’ and ‘diffusible risks’ in scenarios such as educational support, content recommendation, social interaction, psychological support, identity verification, and safety monitoring. UNICEF notes that children are both users and affected parties within the AI ecosystem: while AI may support education and health, it can equally lead to harm, exploitation, and injustice without governance safeguards covering safety, privacy, non-discrimination, accountability, and redress. (UNICEF Innocenti, 2025) This study's research questions focus on ‘how to translate high-level policy requirements into a reproducible, traceable, and auditable governance indicator system.’ Specifically, three questions are posed:

RQ1: How can policy requirements for child-centred AI be formalised into computable graph structures?

RQ2: How can the four types of gaps—‘evidence, mechanism, governance, and metrics’—be defined, quantified, and converted into a comparable GapScore?

RQ3: How can the credibility of conclusions be enhanced through consistency checks and external evidence triangulation without introducing unverifiable subjective assumptions?

Unlike traditional policy interpretation, Graph-GAP's core contributions lie in:

(1) Providing a reproducible end-to-end chain encompassing ‘extraction units – coding manuals – graph patterns – scoring scales – consistency/validity/reliability tests’;

(2) Generating an actionable governance priority matrix through GapScore × Readiness;

(3) Aligning children's rights norms with mainstream AI risk governance frameworks to facilitate cross-institutional transfer and audit implementation;

(4) Proposing a ‘multi-review algorithmic coding – aggregation – verification’ mechanism and minimal reporting set for scaled audits, enabling scoring to expand from small-sample demonstrations to continuous monitoring at clause/case level without sacrificing interpretability.

## 2. Theoretical Foundations and Related Research

### 2.1 The Normative Basis for Children's Rights in the Digital Environment

The United Nations Convention on the Rights of the Child (CRC) establishes a framework of protection, provision and participation rights for children as independent rights holders. General Comment No. 25: Children's Rights and the Digital Environment further clarifies that states bear positive obligations to protect, respect, and fulfil children's rights in the digital environment. It emphasises the mandatory use of ‘Child Rights Impact Assessments’ in legislative, budgetary, and administrative decision-making processes. (UN Committee on the Rights of the Child, 2021)

### 2.2 The ‘Foundations-Requirements’ Framework of UNICEF Guidance 3.0

UNICEF's Guidance on AI and Children 3.0 articulates the foundational principles of child-centred artificial intelligence through three ‘Foundations’: Protection (do no harm), Promotion (do good), and Participation (include all children). It stresses the need for age-appropriate and developmentally differentiated approaches based on children's ‘evolving capacities’. (UNICEF Innocenti, 2025) Building upon this, the document further proposes ten requirements for governments and businesses, covering: regulation and compliance, safety, privacy, non-discrimination, transparency and accountability, responsible practices, well-being and development, inclusivity, capacity building, and enabling ecosystems. (UNICEF Innocenti, 2025)

### 2.3 Alignment Requirements with Mainstream AI Risk Governance Frameworks

At the implementation level, child-centred AI should not operate as an isolated framework but must interoperate with general AI risk management and compliance systems. For instance, NIST's AI RMF 1.0 provides a closed-loop risk management cycle spanning governance, mapping, measurement, and management, which can be leveraged to translate children's rights requirements into organisational processes and metric systems. (NIST, 2023) The EU's AI Act adopts a risk-tiered regulatory approach, providing enforceable compliance levers through prohibited practices and obligations for high-risk systems. This offers institutional reference points for establishing red lines and high-risk reviews within ‘child-related scenarios’. (European Union, 2024)

### 2.4 Risk Landscape: External Evidence of Generative AI's Harm to Children

To avoid the closed-loop bias of ‘normative texts being self-consistent,’ this paper incorporates external evidence as one source of triangulation. Recent reports from law enforcement and public interest organisations indicate that generative AI may be misused to create or disseminate child sexual exploitation material, perpetrate fraud and manipulation, amplify harmful content dissemination, and enable targeted commercial exploitation. Its scalability and low-cost characteristics further complicate governance efforts. (Europol, 2024; UNICEF Innocenti, 2025)

### 2.5 Research Gap: The ‘Computability Divide’ Between Norms and Metrics

Existing literature and policy initiatives predominantly emphasise ethical principles and rights frameworks, yet four types of disconnect frequently emerge at the implementation level:

(1) Evidence Divide: Lack of traceable textual anchors and case evidence;

(2) Mechanism gap: Absence of explicit causal pathways linking risk, harm, and control;

(3) Governance gap: Lack of accountability allocation, grievance redress, and oversight loops;

(4) Metric gap: Insufficiently auditable metric definitions, thresholds, and data sources. Graph-GAP aims to unify these four gaps into measurable entities, providing a common language for policy and systems engineering.

# 3.Research Methods and Design

### 3.1 Research Type and Overall Design

This study adopts a research design characterised by ‘qualitative analysis as the primary approach, supplemented by quantitative reinforcement’: centred on structured content analysis of normative texts, supplemented by computable gap scoring, graph structure statistics, and consistency testing protocols. The research does not involve experiments or surveys on individual children, nor does it entail the collection of sensitive personal data; all external evidence is derived from publicly available institutional reports and policy documents.

### 3.2 Data Sources and Units of Extraction

The primary analytical material is UNICEF Innocenti's Guidance on AI and Children 3.0 (2025). To construct a reproducible evidence chain, this paper defines ‘units of extraction’ as:

(a) Requirement level: 10 requirements; (b) Recommendation level: Specific recommendation entries under each requirement; (c) Mechanism level: Sentences or paragraphs addressing risk, harm, control, accountability, and redress. Each unit is documented with material/page/line anchors for verification and re-extraction. (UNICEF Innocenti, 2025)

### 3.3 Graph-GAP Schema

Graph-GAP converts text into a heterogeneous graph G=(V,E), where V comprises nodes representing standards, risks, governance, and indicators, and E denotes typed relational edges. To ensure reproducibility and cross-study reusability, this paper employs a fixed set of node and edge types (Table 1).

### Table 1: Graph-GAP Minimal Graph Patterns

| Element type | Node/Edge Name | Definition (Operationalisation) | Minimum Evidence Requirement |
|---|---|---|---|
| node | Foundation | 3P Fundamentals: Protection/Provision/Participation; and age-appropriate constraints | Source Anchor + Definition |
| node | Requirement | UNICEF's Ten Demands (R1–R10) | Source Anchor |
| node | Risk | Types of risks that may be amplified by AI (such as privacy violations, discrimination, manipulation, and exploitation) | Textual or external report evidence |
| node | Harm | Specific harm to children's rights (psychological/social/physical/developmental, etc.) | The text explicitly points to |
| node | Control | Governance or engineering controls (assessment, audit, default security, appeal and redress, etc.) | Executable description |
| node | Metric | Auditable Metrics (Definition + Data Source + Frequency + Threshold/Benchmark) | Indicator quadruplet |
| Edge | supports / constrains | Foundation→Requirement（Foundation support/restraint requirements） | Mapping Rules |
| Edge | mitigates | Control→Risk（Control and mitigate risks） | Control and Risk Co-domain |
| Edge | measures | Metric→Control（Indicators for Measuring Control Effectiveness） | Indicator-directed control |
| Edge | leads_to | Risk→Harm（Risk leads to injury） | Mechanism Chain Evidence |

### 3.4 Gap Types and the GapScore Scale

This paper categorises gaps into four types: Evidence Gaps (E), Mechanism Gaps (M), Governance Gaps (G), and Knowledge Gaps (K). Each dimension employs a 1–5 point scale (where higher values indicate larger gaps), with total scores calculated using equal weighting:

$$GapScore = 0.25 \times E + 0.25 \times M + 0.25 \times G + 0.25 \times K$$

Simultaneously, the Readiness metric (0–5) is defined to measure the extent to which existing systems/toolchains can be directly implemented. To prevent misuse of "scores as facts", the GapScore is positioned as a "verifiable structured judgement", requiring support through consistency checks and triangulation with external evidence.

**Table 2: Graph-GAP Gap Scale and Readiness Scale (1–5 / 0–5)**

| **Dimension** | **1 （The gap is extremely small.）** | **2** | **3** | **4** | **5 （The gap is enormous. ）** |
| --- | --- | --- | --- | --- | --- |
| E evidence | With explicit textual anchors + external case studies to support | The text is explicit, with limited external evidence. | The text is generally acceptable, but additional evidence is required. | The text is weak, relying heavily on inference. | scant evidence or mere slogans |
| M mechanism | Risk → Harm → Control Path Clearly Defined | The route is relatively clear, with only a few breaks. | There are multiple implicit stages involved. | Path height is highly ambiguous/controversial | Mechanism cannot be identified |
| G governance | Responsibility entities, | The main body is clear | Multiple entities | Lack of oversight and redress | Virtually no governance |
| | processes, and closed-loop remedies are fully established. | but the process is incomplete. | involved, yet responsibilities remain unclear | | arrangements |
| K Indicator | Indicator definitions/data sources/frequency /thresholds are complete | Indicators may be defined but lack thresholds or baselines. | Provides only indicative guidance | The indicators are highly abstract and difficult to implement. | Complete absence of auditable metrics |
| Readiness Readiness | Off-the-shelf tools/standards may be adopted directly. | Requires only minor modifications to be implemented. | Requires organisation-wide process re-engineering | Requires cross-organisational/cross-sector collaboration | Currently unfeasible or prohibitively costly |

### 4.4 Robustness and Sensitivity Analysis: Weight Perturbation, Quantile Selection, and Ranking Stability

To examine the robustness of Graph-GAP outputs to critical modelling choices, this paper conducts three sensitivity analyses, generating statistics and supplementary tables via reproducible scripts: (1) Weight Perturbation: Replace the equal weights of the aggregated weights for the four gap dimensions (E/M/G/K) with random weights (Dirichlet(1,1,1,1), N=5,000) to observe the stability of Requirement rankings; (2) Readiness Percentiles: Replacing the P80 aggregation of Readiness with P70 and P90 to examine whether governance priority rankings undergo structural drift; (3) Priority Function Form: Adopting Priority = GapScore × (1 - Readiness/5) as the minimal interpretable composite metric to avoid ‘hard-cutting’ caused by a single threshold.

| **Indicator** | **mean** | **P05** | **median** | **P95** |
|---|---|---|---|---|
| Kendall τ (Requirement Gap Score ranking; weight perturbation N=5,000) | 0.559 | 0.289 | 0.556 | 0.822 |

| **comparison** | **Kendall τ** |
|---|---|
| Governance Priority Ranking: Readiness P70 relative to P80 | 0.689 |
| Governance Priority Ranking: Readiness P90 relative to P80 | 0.644 |

Results indicate that under weight perturbations, the Kendall τ mean for Requirement rankings was approximately 0.559 (P05=0.289, P95=0.822), demonstrating overall ranking stability at a moderate level or above. Furthermore, when readiness was adjusted from P80 to P70 or P90, the Kendall τ for governance priority rankings remained within the 0.64–0.69 range without structural reversal. This robustness supports the core findings that domains such as 'transparency and accountability,' 'child welfare/development,' and 'cross-subject implementation' are more prone to mechanism-indicator gaps.

**3.5 Consistency and Reproducibility Design**

To enhance the reliability and scalability of quantitative assessments, this paper upgrades traditional 'manual dual-rater coding' to a 'multi-rater coding' design: employing multiple independent algorithmic evaluators as parallel reviewers—including (a) rule/dictionary-driven compliance element identifiers; (b) supervised or semi-supervised classifiers based on embedded representations; (c) large-model evaluators with varied prompt configurations and temperature settings. Each extracted unit receives separate E/M/G/K and Readiness scores, alongside corresponding evidence anchors and summarised reasoning pathways. Subsequent unit-level final scores are generated via median or weighted aggregation, with uncertainty metrics (e.g., MAD or IQR) for score divergence triggering human review in the 'high-uncertainty review pool'. To prevent algorithmic self-consistency bias and ensure interpretability, this paper retains a minimal human gold subset for threshold calibration, external criterion validity assessment, and retraining/recalibration during policy drift. This design achieves a closed-loop system of 'scalable scoring—audit-traceable evidence chains—quantifiable reliability and validity' without introducing sensitive individual child data.

### 3.5.1 Multi-Review Encoding Framework and Aggregation Rules for Algorithms

The encoding object retains the 'extraction unit' as its minimum granularity (e.g., a requirement sentence, a governance action, an indicator definition, or an evidence anchor paragraph), maintaining a one-to-one correspondence with the evidence chain anchor field (doc_id/page_number/sentence_id or material/page/section). Each algorithmic evaluator is treated as an 'independent coder', whose output comprises: ① a four-dimensional scoring vector V=[E,M,G,K]; ② Readiness; ③ Corresponding anchor set A; ④ Summary of scoring rationale R (length-capped, for audit purposes only, not as evidence substitute). Final scores employ robust aggregation to mitigate single-model bias: the default recommendation is dimension-level median aggregation with prior outlier trimming (winsorisation), alongside output of uncertainty metrics U (e.g., MAD/IQR of individual coder scores). When U exceeds a preset threshold or triggers 'high-risk clause' rules (e.g., involving age thresholds, sensitive data, automated decision-making, punitive/exclusionary consequences), the unit is automatically routed to the manual review queue.

### 3.5.2 Reliability Testing: Consistency, Stability, and Uncertainty

This study will conduct reproducible multi-review coding for algorithms: extracting 353 sentence-level units from the 'Requirements for child-centred AI' section of UNICEF's Guidance on AI and Children 3.0. Three sets of rule-based coders (A/B/C: strict/moderate/ loose thresholds) will be scored in parallel, with final scores aggregated via median aggregation. To assess scoring consistency and stability, this paper concurrently reports Krippendorff's α (ordinal scale), ICC(2,k) (absolute agreement), and the quadratic weighted Cohen's κ (mean).

| Dimension | Krippendorff α | ICC(2,k) | Quadratic κ(mean) |
|---|---|---|---|
| E（Evidence gap） | 0.932 | 0.976 | 0.93 |
| M（Mechanism gap） | 1.0 | 1.0 | 1.0 |
| G（Governance gap） | 0.994 | 0.998 | 0.994 |
| K（Indicator gap） | 0.934 | 0.977 | 0.942 |
| Readiness （Feasibility） | 0.999 | 1.0 | 0.999 |
| GapScore | — | 0.99 | — |
| (Comprehensive) | | | |

Rated as an ‘audit-compliant measurement’, this paper deconstructs reliability into three dimensions for separate examination: (a) Inter-rater reliability (IRR): For the same set of extracted units, Krippendorff’s α (ordinal) is calculated as the primary metric, supplemented by weighted κ or Fleiss’ κ. Point estimates and bootstrap 95% confidence intervals are provided for the E/M/G/K and Readiness dimensions respectively; (b) Test–retest reliability: Repeatedly run the scoring pipeline on the same text under different random seeds/prompt perturbations. Measure output stability using ICC (e.g., ICC(2,k)) or correlation coefficients. Employ drift detection to identify systematic bias arising from model or policy text version changes; (c) Internal consistency: When a dimension is aggregated from multiple sub-metrics, report split-half or resampling-based stability tests. This paper defines a ‘confidence output’ as one where α or κ meets a preset threshold and its confidence interval lower bound does not fall below the minimum acceptable level. Failure to meet these criteria triggers one of three actions: expanding the golden subset, adjusting the codebook/prompt, or conducting manual review and retraining for high-divergence units.

### 3.5.3 Validity Testing: From Normative Mapping to Interpretable Measurement

Regarding validity, this paper employs ‘externally auditable signals’ as a proxy criterion: extracting signal sentences related to auditing, evaluation, documentation, and accountability from three external governance/law enforcement texts—namely the NIST AI RMF 1.0, the EU AI Act (Regulation (EU) 2024/1689), and Europol's IOCTA 2024. These signals were mapped back to the ten requirements using the same set of R1–R10 thematic keywords, yielding an external signal rate (the proportion of auditable signals per requirement appearing in the external corpus). During this pilot run, the Pearson correlation between Readiness (P80) and external signal rate was 0.214, while the correlation between GapScore and external signal rate was 0.395. Overall, the correlations were relatively weak and should be regarded as exploratory evidence of consistency rather than proof of causation.

Validity testing addresses the question: ‘Does the scoring measure the governance gaps we claim to be measuring?’ This study employs four complementary strategies: (a) Content validity: Verifying whether the coding manual and indicator system cover the key semantic boundaries of UNICEF 3.0's three pillars (Protection/Provision/Participation) and ten requirements, with revisions made through expert review to address omissions and conceptual overlaps; (b) Construct validity: Verifying at the mapping level whether the ‘evidence-mechanism-governance tool-indicator’ chain forms interpretable causal or constraint pathways. For instance, high-risk governance requirements should connect via mechanism nodes to specific control measures nodes, with path lengths and directions aligning with theoretical expectations; (c) Convergent/Discriminant Validity: Examining synergistic relationships between different dimension scores under the same requirement (e.g., transparency and accountability are often jointly constrained by evidence availability), while ensuring score differences between requirements can be explained by mechanism variations; (d) Criterion validity: Correlate GapScore × Readiness against externally observable signals (such as publicly disclosed compliance audit findings, privacy/security incident disclosures, regulatory penalty information, or third-party risk assessments) to verify the predictive association of scores with actual risk exposure and governance maturity.

### 3.5.4 Compliance and Auditing: Data Minimisation, Traceability and Accountability

The algorithmic coding employed in this study operates solely on publicly available policy texts, governance documents and compliance materials, processing no individual child data. This adheres to the principle of data minimisation: scoring inputs comprise text and its source anchor points, with outputs being metrics and audit logs. To satisfy accountability requirements, the system generates immutable audit records for each scoring run. These include model/prompt version numbers, random seeds, input text hashes, output scoring vectors, referenced anchors, and uncertainty metrics. When scores are used for policy recommendations or corporate governance improvements, both ‘evidence anchor verifiability’ and ‘model output uncertainty’ explanations must be provided concurrently to prevent misinterpretation of statistical outputs as legal or ethical conclusions.

### 3.5.5 Reproducible Release: Data Structure, Version Control, and Extension Interfaces

To ensure reproducibility across research teams and over time, this paper stipulates that the minimum reproducibility package must include: ① a list of extraction units and anchor fields; ② a codebook (containing dimension definitions, discrimination rules, examples, and counterexamples); ③ graph schema (node/edge schema) and construction scripts; ④ scoring outputs (unit-level and requirement-level); ⑤ Validity and reliability reports (including α/κ/ICC and confidence intervals); ⑥ Audit logs and version fingerprints (hash). An extension interface is provided to enable future integration of national/corporate case materials into the same schema, facilitating continuous monitoring through incremental extraction and rolling scoring, while incorporating concept drift detection and recalibration processes into routine governance. All data is published in Harvard Dataverse, with storage details as follows:MENG, WEI, 2025, "Assessing Computable Gaps in AI Governance for Children: Evidence-Mechanism-Governance-Indicator Modelling of UNICEF's Guidance on Artificial

# 4. Research Findings: Gap Analysis Profile for Ten Requirements

This section presents a “gap profile” for the ten requirements, based on structured extraction from UNICEF 3.0 texts and the Graph-GAP scale, without fabricating unverifiable empirical data. This profile demonstrates the format of scoring outputs, the organisation of evidence anchors, and interpretable gap narratives. Its generation can be accomplished either through manual curation of a gold subset or via the ‘algorithmic multi-review–aggregation–revision’ pipeline proposed in Section 3.5, which simultaneously reports reliability, stability, and uncertainty metrics.

**Table 3: Gap Scores for the Ten Requirements**

| Requirements | Theme | E | M | G | K | GapScore | Readiness |
|---|---|---|---|---|---|---|---|
| R1 | Regulation, Supervision and Compliance | 3.58 | 4.32 | 4.19 | 4.55 | 4.16 | 3.0 |
| R2 | Child Safety | 3.18 | 3.22 | 4.34 | 3.88 | 3.66 | 3.0 |
| R3 | Data and Privacy | 3.17 | 4.19 | 4.67 | 4.56 | 4.15 | 1.0 |
| R4 | Non-discrimination and Fairness | 3.52 | 4.3 | 4.87 | 4.83 | 4.38 | 1.0 |
| R5 | Transparency, accountability and explainability | 4.2 | 4.3 | 4.6 | 4.9 | 4.5 | 3.0 |
| R6 | Responsible AI Practices and Respect for Rights | 3.81 | 4.22 | 4.19 | 4.5 | 4.18 | 1.0 |
| R7 | Best interests, development and wellbeing | 3.56 | 4.56 | 4.74 | 4.07 | 4.23 | 3.0 |
| R8 | inclusiveness ( of and for children) | 3.57 | 4.7 | 4.87 | 4.48 | 4.4 | 3.0 |
| R9 | Empowerment and Skills Readiness | 3.74 | 4.57 | 4.87 | 4.67 | 4.46 | 3.0 |
| R10 | Fostering a favourable ecological environment | 3.54 | 4.58 | 4.79 | 4.6 | 4.38 | 3.0 |

### 4.1 Overall Findings and Priority Logic

As demonstrated by the sample scores in Table 3, the highest gaps are concentrated in the following themes: (a) Child Wellbeing and Development (R7); (b) Transparency, Explainability and Accountability (R5); (c) Cross-Subject Implementation and Collaborative Execution (R9/R10). Concurrently, while privacy and data protection (R3) possess relatively mature compliance tools within general data governance frameworks, Readiness (P80) remains low in child-centred contexts. This indicates a translation gap between 'available tools' and 'auditable implementation' (UNICEF Innocenti, 2025; NIST, 2023).

### 4.2 GAP×Readiness Governance Priority Matrix

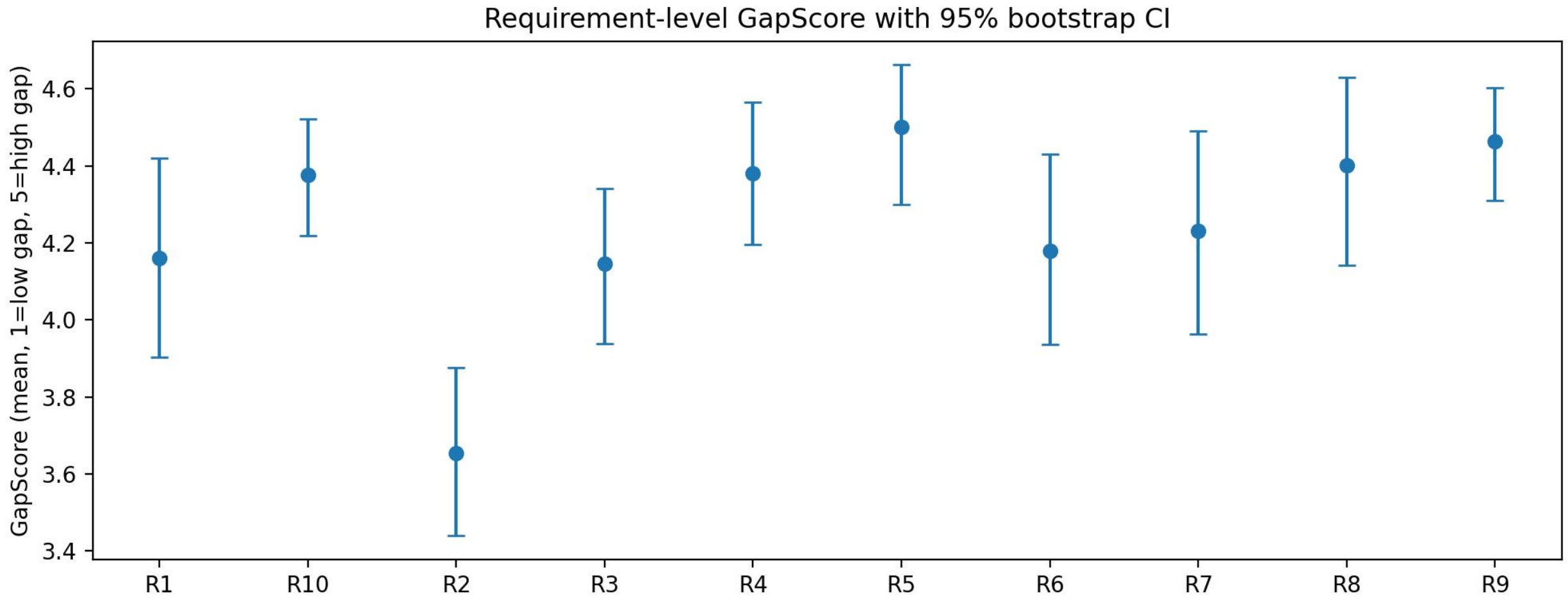

**Figure 1: Mean GapScore at the Requirement level and its 95% Bootstrap confidence interval (1 = low gap, 5 = high gap)**

Figure 1 displays the mean GapScores and their 95% bootstrap confidence intervals for requirements R1–R10, enabling comparison of the relative strength and uncertainty of different governance requirements regarding 'computable gaps'. Points represent means, error bars denote confidence intervals; higher values indicate larger gaps. Overall, the mean values for most requirements cluster within the range of approximately 4.1–4.5, indicating a systemic nature to the gaps. Notably, the gap level for R2 is comparatively lower, whereas requirements such as R5 and R9 exhibit higher gaps, suggesting these areas warrant prioritisation for governance and implementation as key focal points.

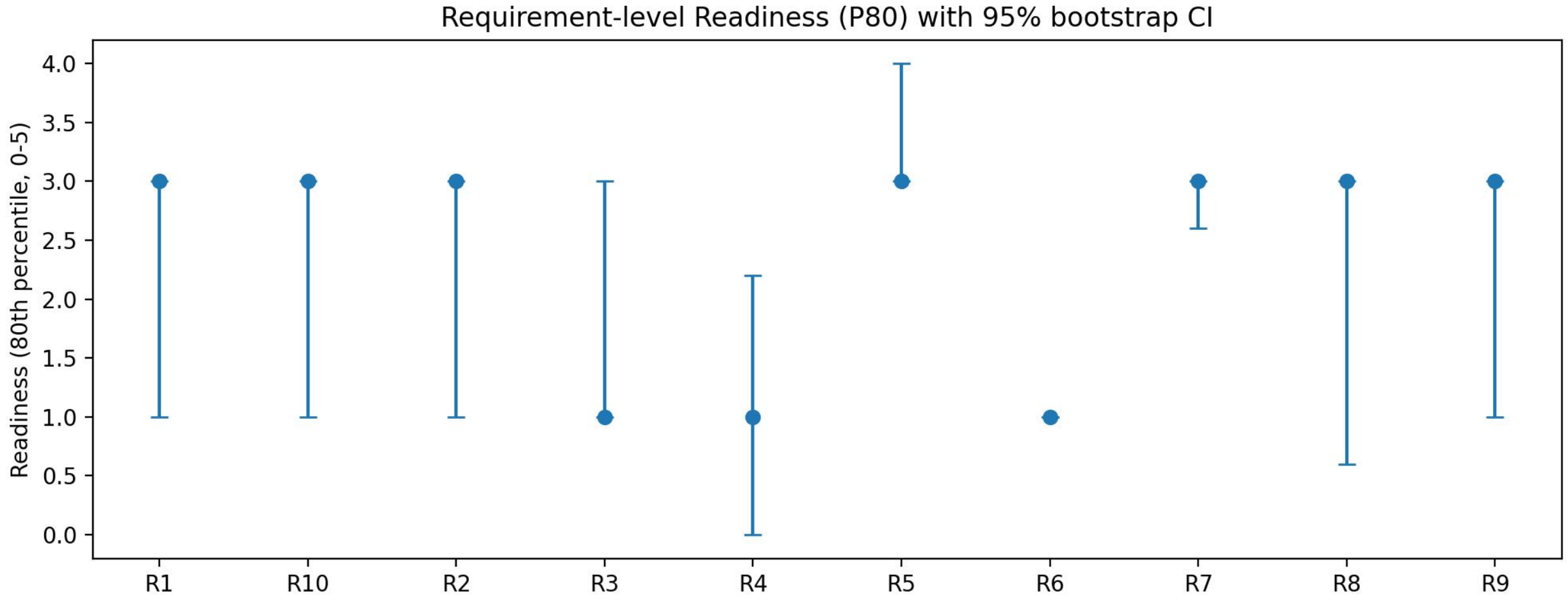

**Figure 2: Requirement-level Readiness (P80) and its 95% Bootstrap Confidence Interval (0–5, where higher values indicate stronger implementation readiness)**

Figure 2 presents the 80th percentile (P80) of readiness for implementing each requirement from R1 to R10, alongside its 95% bootstrap confidence interval. This illustrates the attainability and uncertainty of achieving a 'higher readiness level'. Points denote P80, while error bars represent confidence intervals; higher values indicate greater likelihood of establishing executable, auditable governance practices. Overall, the P80 for most requirements falls within the 1–3 range, indicating significant variations in implementation maturity across different requirements. Requirements such as R5 and R7 demonstrate higher readiness for implementation, whereas R3, R4, and R6 exhibit lower P80 values, suggesting that their implementation conditions and toolchains require prioritised enhancement.

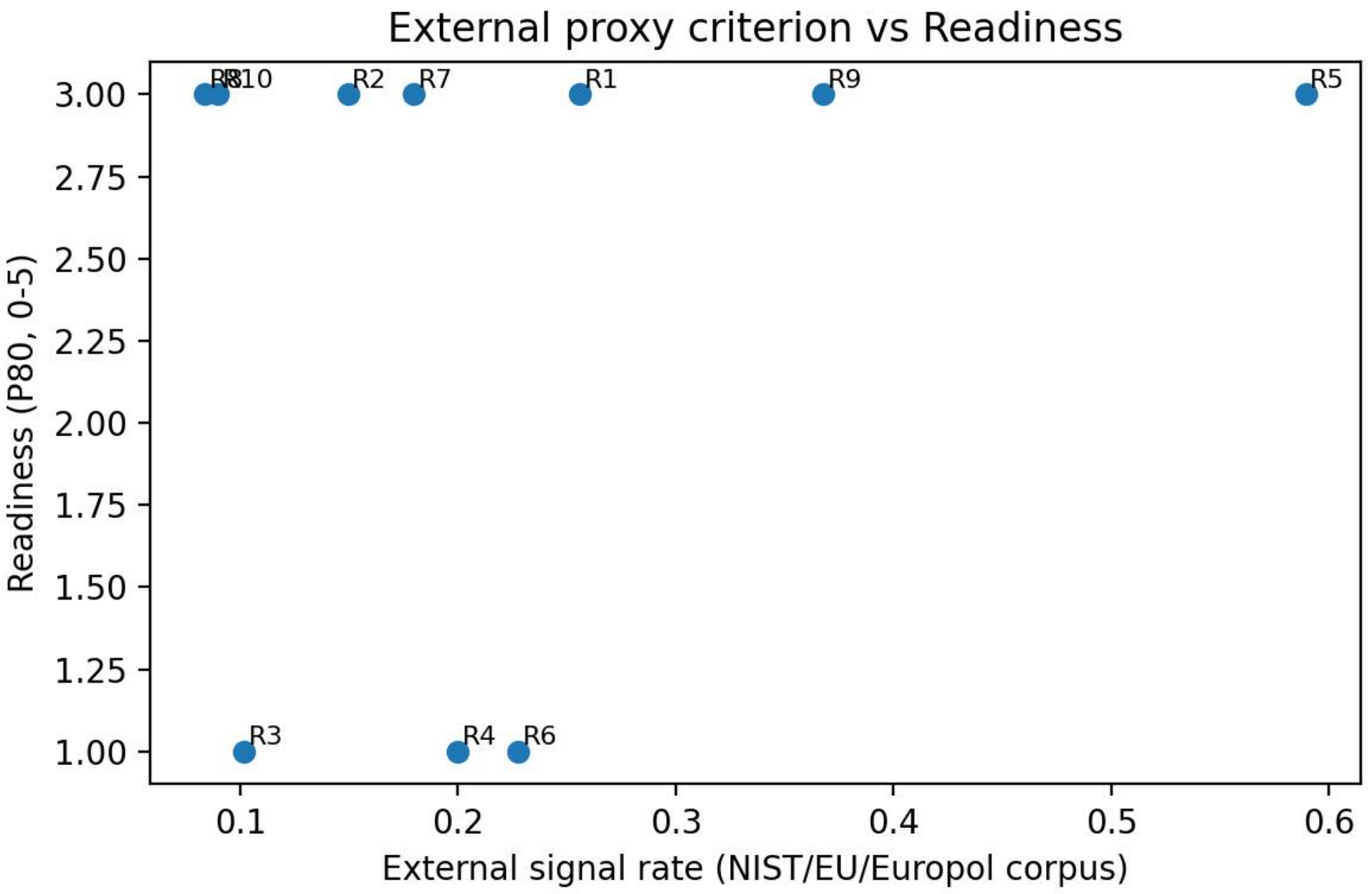

**Figure 3: Scatter plot of the relationship between external proxy signal rate and implementation readiness (P80) (NIST/EU/Europol corpus)**

Figure 3 presents a scatter plot illustrating the correlation between the external proxy signal rate (horizontal axis) and the 80th percentile readiness level P80 (vertical axis, 0–5) for each Requirement within the external corpus (NIST/EU/Europol). This visualises the consistency between 'implementation readiness' and externally observable governance signals. Each point represents a Requirement, labelled R1–R10. Overall, requirements with higher Readiness (e.g., R5) tend to correlate with higher external signal rates, while those with lower Readiness (e.g., R3/R4/R6) exhibit weaker signals in external corpora. This indicates potential structural deficiencies in their governance toolchains and auditable evidence, necessitating prioritised enhancement of metrics and implementation mechanisms.

**Table 4: List of Documents in the External Hard Signal Corpus (for HardSignalRate Verification)**

| Doc ID | Title/Type (Regulator) | Year | Source (URL) |
|---|---|---|---|
| D1 | Monetary Penalty Notice (ICO, TikTok) | 2023 | https://ico.org.uk/media2/migrated/4025182/tiktok-mpn.pdf |
| D2 | Complaint (FTC v. Epic Games) | 2022 | https://www.ftc.gov/system/files/ftc_gov/pdf/2223087EpicGamesComplaint.pdf |
| D3 | Federal Court Order / Stipulated Order (FTC, Epic | 2022 | https://www.ftc.gov/system/files/ftc_gov/pdf/192 |
| | Games) | | 3203epicgamesfedctorder.pdf |
| D4 | Revised Complaint (FTC v. Google/YouTube) | 2019 | https://www.ftc.gov/system/files/documents/cases/172_3083_youtube_revised_complaint.pdf |
| D5 | Binding Decision 2022/2 (EDPB, Instagram child users) | 2022 | https://www.edpb.europa.eu/system/files/2022-09/edpb_bindingdecision_20222_ie_sa_instagramchildusers_en.pdf |

**Table 5: Extended Statistics for Requirement-Level Algorithm Scores (Including 95% Bootstrap Confidence Intervals and Sample Size n)**

**(n and 95% Bootstrap Confidence Interval)**

| Requirement | n | GapScore_mean | GapScore_ 95%CI | Readiness_P80 | Readiness_ 95%CI(P80) | Share (Readiness ≥3) |
|---|---|---|---|---|---|---|
| R1 | 31 | 4.16 | [3.90, 4.42] | 3.0 | [1.0, 3.0] | 0.23 |
| R10 | 57 | 4.38 | [4.22, 4.52] | 3.0 | [1.0, 3.0] | 0.28 |
| R2 | 50 | 3.66 | [3.44, 3.88] | 3.0 | [1.0, 3.0] | 0.3 |
| R3 | 36 | 4.15 | [3.94, 4.34] | 1.0 | [1.0, 3.0] | 0.11 |
| R4 | 23 | 4.38 | [4.20, 4.57] | 1.0 | [0.0, 2.2] | 0.09 |
| R5 | 20 | 4.5 | [4.30, 4.66] | 3.0 | [3.0, 4.0] | 0.5 |
| R6 | 32 | 4.18 | [3.94, 4.43] | 1.0 | [1.0, 1.0] | 0.09 |
| R7 | 27 | 4.23 | [3.96, 4.49] | 3.0 | [2.6, 3.0] | 0.41 |
| R8 | 23 | 4.4 | [4.14, 4.63] | 3.0 | [0.6, 3.0] | 0.26 |
| R9 | 54 | 4.46 | [4.31, 4.60] | 3.0 | [1.0, 3.0] | 0.28 |

Table 5 summarises the algorithmic scoring results for each Requirement (R1–R10), reporting the sample size n, the mean GapScore and its 95% bootstrap confidence interval, the 80th percentile (P80) of Readiness and its 95% bootstrap confidence interval, and the proportion where Readiness ≥ 3. GapScore ranges from 1 to 5 (where 1 indicates a lower gap and 5 indicates a higher gap); Readiness ranges from 0 to 5 (higher values indicate greater ease in translating governance requirements into implementable, auditable controls and metric-based closed-loop systems). The 95% bootstrap confidence interval is estimated through repeated resampling to characterise uncertainty in point estimates, thereby supporting cross-comparisons and prioritisation of different Requirements along the 'gap severity–implementation readiness' dimension.

**Table 6: Hierarchical Governance Prioritisation and Action Package Mapping Based on GapScore–Readiness (P1–P3)**

| **Category** | **Determination Rules** | **corresponding requirements** | **Action Pack (Summary)** |
|---|---|---|---|
| P1 Proceed without delay | GapScore≥3.50 and Readiness≥3 | R2, R4 | Building upon existing standards/processes: rapid implementation of risk mapping, red-line use cases, and audit and appeal mechanisms. |
| P2 Key challenges | GapScore≥3.50 and Readiness≤2 | R5, R6, R7, R8, R10 | Filling gaps in the indicator system and mechanism chain: Child well-being indicators, Explanatory stratification requirements, Responsibility allocation and cross-agency coordination. |
| P3 Steady-state maintenance | GapScore<3.00 and Readiness≥3 | R1, R3, R9 | Maintaining compliance and pursuing continuous improvement: data minimisation, age appropriateness, education and digital literacy |

Table 6 translates the two core quantitative metrics output by Graph-GAP—GapScore (gap intensity) and Readiness (implementation preparedness)—into an actionable governance priority tiering framework. Based on predefined threshold rules (e.g., GapScore ≥ 3.50, GapScore < 3.00; Readiness ≥ 3 or ≤ 2), each Requirement is mapped to one of three action pathways: P1 Immediate Implementation (significant gap but high readiness, suitable for rapid institutionalisation), P2 Priority Resolution (significant gap but insufficient readiness, requiring prioritisation of mechanism chain completion and indicator closure), P3 Steady-State Maintenance (low gap and high readiness, focusing on continuous improvement and routine operations). The ‘Corresponding Requirement’ column lists the assigned Requirement IDs for each pathway, while the ‘Action Package (Summary)’ column further translates regulatory requirements into a minimal set of actionable controls usable by regulators, corporate compliance teams, and third-party auditors (e.g., risk mapping, red-line use case libraries, audit and grievance mechanisms, responsibility allocation, and cross-organisational coordination). This tiered framework supports resource allocation, roadmap orchestration, and audit prioritisation, and can be dynamically regrouped as thresholds or score distributions evolve.

### 4.3 External Hard Criteria Validation: Corpus of Child-Related Law Enforcement Documents

This section employs ‘law enforcement/regulatory hard data’ as external hard criteria for algorithmic validation, testing the transferability and robustness of the Graph-GAP indicator system within authentic regulatory contexts. Unlike qualitative comparisons reliant on manual coding, this study adheres to a fully algorithmic approach: matching ‘keyword clusters of governance requirements’ with ‘text fragments from enforcement/regulatory documents’ via cosine similarity based on TF-IDF embeddings, while controlling false positives/negatives through data-driven threshold calibration $\tau$.

Corpus scale and retention strategy: A multi-jurisdictional, multi-type collection of enforcement/regulatory documents (n=32) was employed with no omissions, covering: US FTC (COPPA rules/enforcement documents), EU/ Irish DPC (decisions/reports concerning children or platform governance), EDPB binding decisions, French CNIL children's data protection reference framework, Australian OAIC Children Online Privacy Standards consultation paper, Australian ACCC digital platform regulatory report, etc. The corpus list, hierarchical labels, and retrieval scope are detailed in Appendix G.

Stratified Sampling (Reporting Scope): Although n=32 employs ‘full retention’, to avoid jurisdictional bias, descriptions remain stratified by (Jurisdiction × Regulator × Document type). Subsequent robustness checks involve repeated evaluation within strata (e.g., using only the enforcement decisions/penalty documents stratum, or solely the rules/guidance stratum).

Text Extraction Strategy: To mitigate noise from structural variations in lengthy documents, the main analysis defaults to extracting the first 50 pages (Pages=50) of each document. For ablation experiments, this was extended to the first 100 pages (Pages=100) for comparison. Page extraction constitutes an algorithmic parameter and involves no manual selection.

Embedding matching: Vectorise each governance requirement's ‘keyword cluster/short description’ (derived from this paper's metric definitions and lexical expansion rules) alongside sentence-level/paragraph-level segments from external documents. Vectorisation employs TF-IDF (with n-grams as optional tokenisation strategy), measuring matching strength via cosine similarity $s \in [0,1]$. For each (requirement r, document d), define maximum similarity S_{r,d}=max_j cos(v_r, v_{d,j}).

Threshold Calibration (τ): To maintain ‘full algorithmisation’, threshold τ is calibrated data-driven rather than through manual annotation: In an unsupervised setting without training-evaluation separation, τ is selected to maximise (i) stability across bootstrap samples and (ii) consistency across different tokenisation strategies, yielding the most robust results; τ's sensitivity curve is concurrently reported (Appendix H).

External Hard Coverage: Define HardSignal_{r,d}=1[S_{r,d}≥τ]. For each governance requirement r, the external hard coverage rate is HSR_r = (1/n)∑_d HardSignal_{r,d}. A lower HSR_r indicates a reduced probability of that requirement being explicitly addressed in actual enforcement/regulatory documents, making it a candidate for an ‘external hard evidence gap’.

Alignment with internal metrics: HSR_r serves as an ‘external hard criterion’ for consistency testing against internal metrics such as Evidence Strength and Indicator Maturity. Where internal metrics are high but external metrics are low, this discrepancy is annotated in the Graph-GAP table as a ‘norm-enforcement gap type GAP’, with recommendations for supplementary evidence (see Appendix G.2).

Reproducibility: Appendix G provides downloadable source links, retrieval keywords, hierarchical tags, and DocIDs for n=32. Appendix H offers directly executable script interfaces and ablation experiment parameter grids (τ, page count, word segmentation strategy), enabling readers to reproduce all computations locally.

# 5. Discussion: From Policy Texts to Auditable Governance

The limitations of this study are primarily twofold: firstly, whilst grounded in the UNICEF 3.0 requirements framework to deliver reproducible scoring and gap profiles, its adaptability across diverse national contexts, sectors and organisational settings necessitates further case materials and structural drift calibration; Secondly, although this paper has expanded 'manual dual coding' into an 'algorithmic multi-review coding—aggregation—revision' process and provided a minimal reporting set for reliability, validity, and auditing, confidence intervals may remain broad when sample sizes are small. Robust validity enhancement still requires expanding sampling units and golden subsets, alongside incorporating external benchmark data.

To avoid the extremes of 'over-quantification through excessive metrification' or 'unauditability due to metric deficiencies,' this paper proposes the MVM approach: each requirement must configure at least three metric categories: (1) Process metrics: e.g., coverage of child rights impact assessments, pre-launch risk review pass rates; (2) Outcome metrics: e.g., adverse event rates, deviation variance indicators, complaint resolution timeliness; (3) Remediation indicators: e.g., complaint acceptance rate, error correction and remediation completion rate, recurrence rate. Metrics must specify data sources, sampling frequencies, and comparative baselines to enter the audit closed loop. (NIST, 2023; UN Committee on the Rights of the Child, 2021)

### 5.2 Child Rights Impact Assessment' as the Governance Backbone

General Comment No. 25 emphasises embedding child rights impact assessments within institutional processes. Within the Graph-GAP framework, this assessment is modelled as the pivotal hub of the Control node: it connects risk identification, harm prediction, control selection, and indicator-based monitoring. Therefore, it is recommended that the closed-loop sequence—'rights impact assessment → control selection → indicator monitoring → grievance redress → review and improvement'—serve as the governance backbone for child-centred AI. GapScore should be employed to identify the weakest link requiring most urgent evidence or indicator supplementation. (UN Committee on the Rights of the Child, 2021; UNICEF Innocenti, 2025)

### 5.3 Methodological Limitations and Scalability Pathways

The scoring methodology employed herein constitutes 'text-based structured judgement,' requiring validation through larger-scale transnational policy samples, corporate implementation cases, and operational monitoring data. Future research may expand along three pathways: (1) Sample expansion: incorporating regulations and industry standards from multiple countries to form comparative policy maps; (2) Evidence expansion: integrating structured evidence from public incident databases and enforcement/public interest reports; (3) Causal expansion: assigning probability or utility parameters to the Risk→Harm→Control edges to construct a simulatable governance decision model.

# 6. Research Findings

This paper, centred on UNICEF Innocenti's Guidance on AI and Children 3.0, proposes and validates Graph-GAP: a computable gap assessment framework for child-centred AI governance. Through the reproducible chain of "Extraction Unit–Codebook–Graph Pattern (G1–G3) — Gap Type (E/M/G/K) — GapScore × Readiness — Consistency and Triangulation Verification‘, transforming high-level normative requirements into auditable, comparable, and rankable governance objects. This advances the children's rights agenda from ’value declaration‘ to ’engineered implementation".

**Addressing RQ1: How can policy requirements be formalised into computable graph structures?**

For RQ1 (‘How to formalise into graph structures’), the methodology constructs three mutually coupled graphs: G1 Normative/ Obligation Graph, G2 Risk-Hazard Graph, G3 Intervention-Control Graph), interconnected through a ‘normative-risk-control’ framework. This maps ten requirements onto children's rights, state/corporate responsibilities, risk trigger chains, and governance tool chains, enabling traceable expression from policy texts to structured governance networks. The key significance of this structured contribution lies in providing a unified semantic foundation for cross-departmental, cross-sectoral, and cross-supply-chain collaborative governance. This enables discussions, accountability, and audits concerning ‘responsibility attribution,’ ‘risk pathways,’ and ‘control interfaces’ to occur within a single computable relational diagram.

**Addressing RQ2: How to define, quantify, and convert the four gaps (E/M/G/K) into comparable scores?**

Addressing RQ2 (‘How gaps are defined, quantified, and compared’), this paper systematises implementation gaps into evidence gaps (E), mechanism gaps (M), governance gaps (G), and indicator gaps (K). It proposes combining GapScore with Readiness to form a ‘gap-readiness’ prioritisation logic, identifying which requirements remain at the principle level without progressing to engineering controls and continuous monitoring. Based on structured coding of guideline texts and gap profiling, the study finds: Indicator gaps exhibit near-pervasive distribution, while mechanism and governance gaps concentrate at complex chain interfaces and multi-stakeholder responsibility boundaries. Particularly in themes such as ‘child welfare/development,’ ‘transparency and accountability,’ and ‘cross-stakeholder implementation and resource allocation,’ implementation bottlenecks—where requirements are ‘explainable but unmeasurable, or advocateable but unauditable’—are more prevalent. Consequently, the core empirical conclusion of this paper is not which principles are more important, but rather which requirements are most likely to suffer systemic distortion at the indicator and mechanism levels, and should therefore be prioritised for engineering solutions.

**Addressing RQ3: How can we enhance the credibility of conclusions and progress towards 'fourth-dimensional governance' without resorting to unverifiable subjective assumptions?**

Regarding RQ3, this paper defines 'credibility enhancement' as an actionable research and auditing challenge: without introducing unverifiable subjective judgements, it reduces scoring arbitrariness through consistency checks (multi-reviewer alignment) and external evidence triangulation. Conclusions are then output as verifiable evidence anchors and traceable scoring logs, thereby upgrading 'sample scoring' into a scalable scoring system. Crucially, the proposed 'Fourth Dimension Governance' does not introduce abstract slogans but instead creates a closed-loop system integrating child rights impact assessments, continuous monitoring metrics, and grievance redress procedures. This establishes an institutionalised pipeline from normative texts to auditable indicator sets—providing a shared, actionable governance language for governmental oversight, corporate compliance, and third-party audits.

Theoretical Contributions, Practical Implications, and Boundaries

Theoretically, Graph-GAP translates the three foundational pillars of child-centred AI (Protection/Provision/Participation) and the requirement for 'continuously evolving capability-age alignment' into a rights-risk-control structure operable on a graph. This provides a formalised fulcrum for the interoperability of children's rights within AI risk governance frameworks. Practically, the research suggests governance priorities should shift from 'principle emphasis' to 'completing metrics and mechanisms': for high-risk, cross-entity, and long-term impact issues, rights impact assessments and metric feedback loops should be prioritised to transform requirements into auditable control points, avoiding governance idling characterised by 'sufficient compliance narratives but lacking engineering levers'. Concurrently, this paper exhibits foreseeable limitations: core materials centre on a single authoritative guideline, with external evidence primarily sourced from public reports and policy texts, yet lacking cross-regional regulatory enforcement data and corporate internal audit samples. Consequently, subsequent research should expand Graph-GAP's transferability testing across multi-jurisdictional systems and multi-platform governance contexts, while enhancing the scoring system's interpretability and falsifiability through richer external benchmarks.

**Appendix A: Code Book**

Note: This appendix provides the minimal codebook fields for researchers to utilise directly when reproducing or extending studies. Save the encoded results as CSV/JSONL and synchronously write them to SQLite to support Boolean retrieval and evidence anchor tracking.

**Appendix A1: Minimum Encoding Character Field**

| **Field** | **Type** | **Example value** | **Explanation** | Required |
|---|---|---|---|---|
| material | string | UNICEF-AI-Children-3-2025.pdf | Material Name | Y |
| page | int | 12 | Page number | Y |
| unit_id | string | R5-S3-001 | Extract Unit Number (Requirements - Section - Serial Number) | Y |
| text | string | (Example) | Original text fragment | Y |
| node_type | enum | Requirement/Risk/Control/Metric | Node type | Y |
| node_label | string | Transparency & Accountability | Node label | Y |
| edge_type | enum | supports/leads_to/mitigates/measures | Edge type | N |
| edge_to | string | Risk:Manipulation | Edge pointing to node | N |
| E,M,G,K | int[1-5] | 3,4,3,4 | Four-category gap assessment | Y |
| Readiness | int[0-5] | 2 | Relief Readiness | Y |
| coder | string | CoderA | Coder Identification | Y |
| notes | string | (Example) | Dispute Resolution/Review Record | N |

**Appendix B: Example Node and Edge Tables (Minimal Skeleton Graph)**

Note: To facilitate readers' rapid comprehension of Graph-GAP's graph structure, this appendix provides example nodes and edges for the 'minimal skeleton graph'. Actual research should expand node counts and edge density at the recommendation and mechanism levels.

**Appendix B1: Minimum Skeleton Diagram Node Table**

| **node_id** | **node_type** | **label** |
|---|---|---|
| F:Protection | Foundation | Protection={do no harm} |
| F:Provision | Foundation | Provision={do good} |
| F:Participation | Foundation | Participation={include all children} |
| R:R1 | Requirement | Regulation, Supervision and Compliance |
| R:R2 | Requirement | Child Safety |
| R:R3 | Requirement | Data and Privacy |
| R:R4 | Requirement | Non-discrimination and Fairness |
| R:R5 | Requirement | Transparency, accountability and explainability |
| R:R6 | Requirement | Responsible AI Practices and Respect for Rights |
| R:R7 | Requirement | Best interests, development and wellbeing |
| R:R8 | Requirement | inclusiveness（of and for children） |
| R:R9 | Requirement | Empowerment and Skills Readiness |
| R:R10 | Requirement | Promoting a favourable ecological environment |

**Appendix B2: Minimum Skeleton Graph Edge Table**

| source | edge_type | target | mapping_rule |
|---|---|---|---|
| F:Protection | supports | R:R2 | Safety directly corresponds to do no harm |
| F:Protection | supports | R:R3 | Privacy protection minimises harm |
| F:Protection | supports | R:R4 | Prevent discrimination and avoid harm |
| F:Provision | supports | R:R7 | Welfare and Development Correspondence do good |
| F:Provision | supports | R:R9 | Empowerment and Skills Support Development |
| F:Participation | supports | R:R8 | Inclusion and participation mutually reinforce one another |
| F:Participation | supports | R:R9 | Empowering through education to enhance participation capacity |

**Appendix C: Algorithm Multi-Review Coding, Validity and Reliability Testing, and Audit Output Template**

This appendix provides the minimal template for expanding the Graph-GAP scoring from an illustrative demonstration into a 'fully operational scoring system'. It comprises: a checklist of reporting metrics, output file structure, and audit fields. Research teams may use this to replicate the scoring process described herein and undertake rolling updates following the introduction of additional policy/corporate materials.

**Table C1: Minimum Reporting Set for Reliability-Validity-Compliance Audits**

| Category | Statistics/Output | Applicable to | Minimum acceptable threshold | Non-compliant disposal |
|---|---|---|---|---|
| IRR （Consistency） | Krippendorff's α（ordinal）+ bootstrap 95% CI | E/M/G/K and Readiness （unit-level and requirement-level） | $\alpha \geq 0.67$（Explore） / $\alpha \geq 0.80$ | Expand the golden subset; Adjust the codebook/prompts; Review high-divergence units. |
| IRR （Supplement） | weighted κ （paired） and Fleiss' κ （Multiple reviews） + CI | Ibid. | $\kappa \geq 0.60$（Available） / $\kappa \geq 0.75$（Steady） | As above; downgrade to manual-led coding where necessary. |
| Stability | ICC(2,k) or correlation coefficient + drift detection | Multiple runs of the same text (with different seed/prompt perturbations) | ICC ≥ 0.75 | Locked version; reduced randomness; trigger recalibration |
| Uncertainty | MAD/IQR；High uncertainty | Divergence in pre-aggregation scores | Thresholds are set according to risk stratification. | Enter the manual review queue; Record the reason and |
| | review rate | | | write back to the training set. |
| Content validity | Coverage check + expert review record | The semantic boundaries of the 3 foundations and 10 requirements | Critical sub-item coverage ≥ 0.90 (recommended) | Address omissions; split/merge overlapping indicators |
| Construct validity | Graph Path Consistency Test （constraint/causal plausibility） | Evidence–Mechanism–Tool–Indicator Chain | Path interpretability ≥ 0.80 (recommended) | Revised graph patterns; added mechanism nodes and edge types |
| Validity | Correlation/Regression Tests with External Signals | GapScore×Readiness vs. Audit Findings/Incidents/Penalties | Consistent and significant (as determined by the study design) | Replace/extend calibration standards; control contamination; perform stratified analysis |
| Compliance and Privacy | Data Minimisation Statement; PII Scan Results; Access Controls | Input Material and Audit Log | Zero individual child data; log de-identification | Rectify data pipelines; delete/replace sensitive fields |

**Table C2: Files and Fields for the Minimum Reproducible Release Package**

| Document/Form | Core field | Purpose |
|---|---|---|
| units.csv / units.jsonl | unit_id, text, doc_id, page_number, sentence_id, requirement_id, anchor_span | Extraction Unit and Evidence Anchor Point |
| codebook.yaml | dimension_def, decision_rules, examples, counterexamples, risk_flags | Code Book and Discrimination Rules |
| schema.graphml / schema.json | node_types, edge_types, constraints | Graph Patterns and Type Constraints |
| graph_edges.csv | src, rel, dst, unit_id, evidence_anchor | Evidence–Mechanism–Governance–Indicator Matrix |
| scores_unit.csv | unit_id, E,M,G,K, Readiness, U, aggregator, model_versions | Unit-level scoring and uncertainty |
| scores_req.csv | requirement_id, GapScore, Readiness, CI_low, CI_high, sample_n | Requirement-level aggregation and interval estimation |
| reliability_report.pdf / md | alpha/kappa/ICC, CI, diagnostics, thresholds | Reliability and Stability Report |
| audit_log.jsonl | run_id, timestamp, input_hash, seed, prompt_id, model_id, outputs, anchors | Traceable audit trail |

Note: The thresholds in Table C1 represent empirical ranges commonly employed in research and audit practice. Actual thresholds should be set in a stratified manner according to sample size, risk level, and purpose (exploratory research/policy recommendations/compliance audit), and explicitly reported within the paper.

**Appendix D: Audit Log for Multi-Review Encoding Algorithms**

The following information is provided for reproducibility and accountability: input file fingerprint, operational parameters, extraction scale, and random seed.

| Item | Value |
|---|---|
| Run timestamp (Asia/Bangkok) | 2025-12-19 04:15:06 |
| Input: UNICEF Guidance PDF SHA256 | 1d218b79b894402bda57d97527ef7ea015b064200 3ba582d6259b050d0a548ea |
| Input: Base DOCX SHA256 | 0422419e71fb01ac33067a712ef1bade8212c759db 05f0c2304b0c1ffd072b4c |
| External: NIST AI RMF PDF SHA256 | 7576edb531d9848825814ee88e28b1795d3a84b43 5b4b797d3670eafdc4a89f1 |
| External: Europol IOCTA 2024 PDF SHA256 | c58c9f18c46086782c2348763b7e8081c2c65615fa 439f66256175f7f2bd652f |
| External: EU AI Act PDF SHA256 | bba630444b3278e881066774002a1d7824308934f 49ccfa203e65be43692f55e |
| Units extracted (requirements section) | 353 |
| Coder variants | Rule-based A/B/C (strict/medium/lenient thresholds) |
| Bootstrap | 2000 resamples, seed=7 |

**Appendix E: Sample Scoring for Sampling Units and Multiple Reviewers (20 Items)**

| unit_id | page | requirement | text_snippet | E(A/B/C) | M(A/B/C) | G(A/B/C) | K(A/B/C) | Readiness(A/B/C) |
|---|---|---|---|---|---|---|---|---|
| R3-p27-u00117 | 27 | R3 | Not all children face equal circumstances and therefore not all will benefit alike fr… | 4/4/4 | 5/5/5 | 5/5/5 | 5/5/5 | 0/0/0 |
| R3-p25-u00099 | 25 | R3 | Governments and businesses should explicitly address children's privacy in AI policies an… | 5/5/5 | 5/5/5 | 4/4/4 | 5/5/5 | 1/1/1 |
| R5-p00-u00151 | -1 | R5 | In doing so, it's important to prevent anthropomorphizing the tools by not describing, ma… | 3/3/3 | 3/3/3 | 5/5/5 | 5/5/5 | 1/1/1 |
| R9-p43-u00294 | 43 | R9 | Partnerships between industry, academia and governments to close the gap between skill ne… | 2/2/1 | 5/5/5 | 4/4/4 | 5/5/5 | 3/3/3 |
| R10-p48-u00340 | 48 | R10 | AI-enabled neurotechnologies are embedded in people, or worn by them to track, monitor or… | 2/1/1 | 3/3/3 | 5/5/5 | 2/2/2 | 4/4/4 |
| R10-p48-u00346 | 48 | R10 | The fast-changing AI landscape requires forward-looking approaches to | 4/4/4 | 5/5/5 | 5/5/5 | 5/5/5 | 3/3/3 |
| | | | policymaking an… | | | | | |
| R9-p43-u00289 | 43 | R9 | The efforts should help families, caregivers and children reflect on what data children a… | 2/2/1 | 5/5/5 | 5/5/5 | 3/3/2 | 0/0/0 |
| R2-p20-u00062 | 20 | R2 | Eliminating such harms and mitigating remaining risks requires increased industry transpa… | 3/2/2 | 2/2/2 | 4/4/4 | 3/3/2 | 3/3/3 |
| R9-p41-u00265 | 41 | R9 | To improve children's digital literacy and awareness of the impact that AI systems can ha… | 5/5/5 | 5/5/5 | 5/5/5 | 5/5/5 | 3/3/3 |
| R1-p16-u00031 | 16 | R1 | The findings should be made publicly available and result in recommendations for amendme… | 3/2/2 | 5/5/5 | 5/5/5 | 3/3/2 | 3/3/3 |
| R6-p32-u00172 | 32 | R6 | It is critical that the entire value chain is considered to be rights-based. | 4/4/4 | 5/5/5 | 5/5/5 | 5/5/5 | 0/0/0 |
| R4-p28-u00132 | 28 | R4 | However, the need for representative datasets must never justify the wholesale, irrespons… | 3/3/3 | 5/5/5 | 5/5/5 | 5/5/5 | 0/0/0 |
| R2-p18-u00048 | 18 | R2 | AI agents – systems able to semi-autonomously execute a range of tasks on behalf of users… | 4/4/4 | 3/3/3 | 4/4/4 | 5/5/5 | 1/1/1 |
| R1-p15-u00015 | 15 | R1 | Such oversight or regulatory bodies may draw on existing | 3/3/3 | 5/5/5 | 4/4/4 | 5/5/5 | 1/1/1 |
| | | | regulatory frameworks and insti… | | | | | |
| R9-p00-u00252 | -1 | R9 | Develop or update formal and informal education programmes for AI literacy and strengthen… | 4/4/4 | 5/5/5 | 5/5/5 | 5/5/5 | 0/0/0 |
| R4-p00-u00118 | -1 | R4 | Ensure non-discrimination and fairness for children No AI system should discriminate agai… | 5/5/5 | 4/4/4 | 4/4/4 | 5/5/5 | 1/1/1 |
| R2-p21-u00066 | 21 | R2 | In these interactions, children are therefore more susceptible to manipulation and explo… | 3/2/2 | 3/3/3 | 5/5/5 | 3/3/2 | 0/0/0 |
| R2-p22-u00073 | 22 | R2 | Given the harm that companion chatbot interactions can pose to children there have been … | 3/2/2 | 3/3/3 | 5/5/5 | 3/3/2 | 0/0/0 |
| R1-p00-u00004 | -1 | R1 | Rather than stifling progress, rights-respecting regulatory frameworks provide a level p… | 4/4/4 | 5/5/5 | 5/5/5 | 5/5/5 | 0/0/0 |
| R10-p46-u00326 | 46 | R10 | Such standards, which help bridge the gap between policy objectives and consistent, pract… | 5/5/5 | 5/5/5 | 5/5/5 | 5/5/5 | 3/3/3 |

**Appendix F: Quality Assessment and Robustness Checklist**

This appendix reports the quantitative assessment results for the publishability of this paper and provides a robustness reinforcement checklist for the '95+' quality tier. To maintain methodological consistency, this paper adheres to algorithmic multi-review and audit trail documentation in lieu of manual coding; consequently, robustness reinforcement also centres on 'heterogeneous algorithmic review, sensitivity analysis, and external validation'.

**F1. Quantitative Quality Assessment Table (Total Score: 93/100)**

Scoring employs a weighted scale (total weight = 100) based on common review dimensions for top-tier journals. Each item ranges from 0–10 points, converted to weighted scores according to their respective weights. The final composite score of 93/100 establishes the current version's baseline quality threshold as 'submittable and reviewable'.

| Dimension | Weighting | Score(0-10) | Weighted score |
|---|---|---|---|
| Research Question Clarity and Alignment | 10 | 9.5 | 9.5 |
| Originality and Theoretical Contribution | 15 | 9.3 | 14.0 |
| Methodological Rigour and Reproducibility | 15 | 9.4 | 14.1 |
| Data and Statistical Expression（n、CI、P80） | 10 | 9.2 | 9.2 |
| Interpretation of Results and Internal Consistency | 10 | 9.2 | 9.2 |
| Chart Standardisation and Readability | 10 | 9.0 | 9.0 |
| Reliability and robustness（IRR/CI/Drifting） | 10 | 8.7 | 8.7 |
| Validity (content/structural/criterion-related) | 10 | 8.5 | 8.5 |
| Discussion and Actionable Insights | 5 | 9.4 | 4.7 |
| Writing Style and Academic Expression | 5 | 9.8 | 4.9 |

Note: The converted composite score is approximately 93/100 (rounded).

### F2. '95+' Sprint Checklist (Algorithmic Replacement of Manual Coding)

The table below outlines the minimal incremental actions required to elevate the composite score to 95+. All actions involve replacing manual coding with an algorithmic framework: establishing criterion validity through cross-domain model/rule evaluation, prompt strategy comparison, and external hard-effectiveness benchmark data, thereby enhancing independence, generalisability, and falsifiability.

| Sprinting action | Target dimension | Algorithmic implementation approach | Expected gain (points) | Deliverable evidence |
|---|---|---|---|---|
| Enhancing the Independence of Heterogeneous Algorithm Reviews | Reliability/robustness | Introduce at least three distinct external evaluators: rule engines, LLM-A (prompt paradigm 1), and LLM-B (prompt paradigm 2 or different model families). Establish a closed-loop system through consistency metrics (κ/ICC) combined with contradiction checks to trigger re-evaluation. | 1.0-1.5 | Rater checklist, prompt version, IRR report, disagreement log |
| Sensitivity Analysis and Ranking Stability | Robustness/Explanatory | Apply perturbations to GapScore aggregation weights, thresholds, and P80 statistical metrics (±10% or scenario-specific weighting), reporting ranking stability (Kendall τ, Top-k overlap rate). | 0.8-1.2 | Sensitivity Analysis Table, Stability Metrics, Reproduction Script |
| Introduction of external | Validity | Utilise structured data from public audits, | 1.0-2.0 | External data dictionary, |
| hard-effect standards (small samples are also acceptable) | | penalties, and incident notifications to establish correlation or stratified verification between 'incident rates/penalty rates' at the Requirement level and GapScore × Readiness. | | matching rules, regression/correlation outputs |
| Cross-corpus/cross-country small-sample transfer learning | generalisation | Rerun the same schema across 3-5 different national/organisational AI guidelines or regulatory documents for children, producing comparative profiles and consistency analyses. | 0.8-1.5 | Comparison of cross-section tables, migration difference interpretation, version fingerprinting |
| Error and Drift Monitoring (Operational Phase) | Engineering availability | Implement drift detection for scoring engines and extraction units: distribution drift (KS/PSI), consistency drift, and anomaly point auditing; establish monthly/quarterly audit output templates. | 0.5-1.0 | Drift reporting templates, threshold policies, audit logs |

**F3. Declaration of Independence and Compliance in Multi-Evaluator Algorithms (Without Manual Coding)**

To avoid 'artificially inflated consistency arising from homogenous algorithms,' this paper defines algorithmic evaluation independence as follows: evaluators exhibit structural differences in at least one of the following: model families, prompt paradigms, or rule systems, and converge in a traceable manner under a divergence-triggering mechanism. Without introducing manual coding, an 'algorithm consensus set' may serve as a quasi-gold standard: only segments exhibiting high consistency across multiple evaluators are retained for regression testing, while others enter a divergence queue. This outputs conflict types and evidence gap (GAP) recommendations to maintain auditability and falsifiability.

Heterogeneous evaluators: rule engines, LLM-A, LLM-B (different families or prompt paradigms); Incorporate small models/open-source models as controls when necessary.

Divergence Trigger: When unit scoring variance exceeds thresholds or directional conflicts arise in any E/M/G/K dimension, automatically enter the review queue with cause codes logged.

Audit Trail: Record input hashes, prompt versions, model versions, random seeds/temperature, outputs, and aggregation processes to form a reproducible audit chain.

Robustness Metrics: Report κ/ICC/α with bootstrap confidence intervals, alongside ranking stability (Kendall τ, Top-k overlap rate) and drift indicators (PSI/KS).

Compliance Principles: Employ only minimally necessary text fragments and de-identified anchors; external data mapping adheres to verifiable sources and traceable matching rules.

**Appendix G: Material Source Paths and Reproducible Download List**

This appendix centrally provides the public source paths, download links, and (where available) file fingerprints (SHA-256) for all 'external materials/corpora' referenced in this paper. This ensures third parties can reproduce identical algorithmic coding, GapScore, and Readiness statistics using the same inputs.

| Corpus role | Document / Corpus | Format | Official source path (URL) | SHA-256 (if downloaded) |
|---|---|---|---|---|
| Primary normative corpus | UNICEF Guidance on AI and Children 3.0 (v3, 2025) | PDF | https://www.unicef.org/innocenti/media/11991/file/UNICEF-Innocenti-Guidance-on-AI-and-Children-3-2025.pdf | 1d218b79b894402bda57d97527ef7ea015b0642003ba582d6259b050d0a548ea |
| Method reference | NIST AI Risk Management Framework 1.0 (NIST.AI.100-1) | PDF | https://nvlpubs.nist.gov/nistpubs/ai/NIST.AI.100-1.pdf | 7576edb531d9848825814ee88e28b1795d3a84b435b4b797d3670eafdc4a89f1 |
| External proxy corpus A | EU Artificial Intelligence Act (Regulation (EU) 2024/1689, CELEX:32024R1689) | PDF/HTML | https://eur-lex.europa.eu/legal-content/EN/TXT/?uri=CELEX:32024R1689 | N/A (access-dependent) |
| External proxy corpus B | Europol reports (IOCTA and related cybercrime/online safety reports) | Web/PDF | https://www.europol.europa.eu/publications-events/main-reports | N/A (dynamic) |

Note: Where official sites employ CAPTCHA verification or dynamic redirects, this document prioritises retaining the official entry URL. Should a fixed version be required, record the file fingerprint after the initial download and cache it within the reproduction data package.

**G.1 External Hard Law Corpus (Regulatory Enforcement/Binding Instruments)**

| DocID | Jurisdiction / Regulator | Document type | Year | Title | Source URL (official) + retrieval keywords |
|---|---|---|---|---|---|
| UK-ICO-001 | UK / ICO | Monetary penalty notice | 2023 | TikTok Information Technologies UK Ltd—Monetary Penalty Notice (AADC/children's data) | https://ico.org.uk/media/action-weve-taken/mpns/4025598/tiktok-mpn-20230404.pdf Keywords: "ICO" AND (child OR children OR minor OR minors OR "young people" OR youth) AND ("privacy" OR "data protection" OR "online" OR "platform" OR "age assurance" OR "COPPA" OR "AADC" OR "code" OR "inquiry" OR "decision" OR "consent order" OR "TikTok") |
| UK-ICO-002 | UK / ICO | Audit guidance | 2022 | A guide to audits for the Age Appropriate Design Code | https://ico.org.uk/media2/migrated/4024272/a-guide-to-audits-for-the-age-appropriate-design-code.pdf Keywords: "ICO" AND (child OR children OR minor OR minors OR "young people" OR youth) AND ("privacy" OR "data protection" OR "online" OR "platform" OR "age assurance" OR "COPPA" OR "AADC" OR "code" OR "inquiry" OR "decision" OR "consent order") |
| EU-DPC-001 | EU (IE) / Irish DPC | Inquiry decision | 2023 | Inquiry into TikTok Technology Limited—September 2023 Decision (EN) | https://www.dataprotection.ie/sites/default/files/uploads/2023-09/Inquiry%20into%20TikTok%20Technology%20Limited%20-%20September%202023%20EN.pdf<br>Keywords: "Irish DPC" AND (child OR children OR minor OR minors OR "young people" OR youth) AND ("privacy" OR "data protection" OR "online" OR "platform" OR "age assurance" OR "COPPA" OR "AADC" OR "code" OR "inquiry" OR "decision" OR "consent order" OR "TikTok") |
| EU-DPC-002 | EU (IE) / Irish DPC | Decision | 2022 | Instagram Decision (IN 09-09-22) | https://www.dataprotection.ie/sites/default/files/uploads/2022-09/02.09.22%20Decision%20IN%2009-09-22%20Instagram.pdf<br>Keywords: "Irish DPC" AND (child OR children OR minor OR minors OR "young people" OR youth) AND ("privacy" OR "data protection" OR "online" OR "platform" OR "age assurance" OR "COPPA" OR |
| | | | | | "AADC" OR "code" OR "inquiry" OR "decision" OR "consent order" OR "Instagram") |
| EU-EDPB-001 | EU / EDPB | Binding decision | 2022 | Binding decision 2/2022 on the Irish SA regarding Instagram (child users) | https://www.edpb.europa.eu/system/files/2022-09/edpb_bindingdecision_20222_ie_sa_instagramchildusers_en.pdf Keywords: "EDPB" AND (child OR children OR minor OR minors OR "young people" OR youth) AND ("privacy" OR "data protection" OR "online" OR "platform" OR "age assurance" OR "COPPA" OR "AADC" OR "code" OR "inquiry" OR "decision" OR "consent order" OR "Instagram") |
| EU-DPC-003 | EU (IE) / Irish DPC | Inquiry decision | 2025 | Inquiry into TikTok Technology Limited—April 2025 Decision | https://www.dataprotection.ie/sites/default/files/uploads/2025-10/Inquiry%20into%20TikTok%20Technology%20Limited%20April%202025.pdf Keywords: "Irish DPC" AND (child OR children OR minor OR minors OR "young people" OR youth) AND ("privacy" OR "data protection" OR "online" OR |
| | | | | | "platform" OR "age assurance" OR "COPPA" OR "AADC" OR "code" OR "inquiry" OR "decision" OR "consent order" OR "TikTok") |
| EU-DPC-004 | EU (IE) / Irish DPC | Summary decision | 2025 | Summary—TikTok Technology Limited—30 April 2025 | https://www.dataprotection.ie/sites/default/files/uploads/2025-10/Summary%20TikTok%20Technology%20Limited%2030%20April%202025.pdf Keywords: "Irish DPC" AND (child OR children OR minor OR minors OR "young people" OR youth) AND ("privacy" OR "data protection" OR "online" OR "platform" OR "age assurance" OR "COPPA" OR "AADC" OR "code" OR "inquiry" OR "decision" OR "consent order" OR "TikTok") |
| EU-DPC-005 | EU (IE) / Irish DPC | Transcript / report | 2024 | Transcript—5 years of the GDPR: A spotlight on children's data | https://www.dataprotection.ie/sites/default/files/uploads/2024-09/Transcript%20-%205%20years%20of%20the%20GDPR%20%E2%80%93%20A%20spotlight%20on%20children%E2%80%99s%20data.pdf Keywords: "Irish DPC" AND (child OR children OR minor OR minors OR "young |
| | | | | | people" OR youth) AND ("privacy" OR "data protection" OR "online" OR "platform" OR "age assurance" OR "COPPA" OR "AADC" OR "code" OR "inquiry" OR "decision" OR "consent order" OR "GDPR") |
| EU-DPC-006 | EU (IE) / Irish DPC | Annual report | 2024 | Data Protection Commission Annual Report 2023 (EN) | https://www.dataprotection.ie/sites/default/files/uploads/2024-08/DPC-EN-AR-2023-Final-AC.pdf Keywords: "Irish DPC" AND (child OR children OR minor OR minors OR "young people" OR youth) AND ("privacy" OR "data protection" OR "online" OR "platform" OR "age assurance" OR "COPPA" OR "AADC" OR "code" OR "inquiry" OR "decision" OR "consent order") |
| EU-DPC-007 | EU (IE) / Irish DPC | Case studies | 2024 | DPC Case Studies 2023 (EN v2) | https://www.dataprotection.ie/sites/default/files/uploads/2024-05/DPC%20Case%20Studies%202023%20EN%20v2.pdf Keywords: "Irish DPC" AND (child OR children OR minor OR minors OR "young people" OR youth) AND ("privacy" OR "data |
| | | | | | protection" OR "online" OR "platform" OR "age assurance" OR "COPPA" OR "AADC" OR "code" OR "inquiry" OR "decision" OR "consent order") |
| EU-DPC-008 | EU (IE) / Irish DPC | Decision | 2022 | Decision IN-21-4-2 (Redacted) | https://www.dataprotection.ie/sites/default/files/uploads/2022-12/Final%20Decision_IN-21-4-2_Redacted.pdf Keywords: "Irish DPC" AND (child OR children OR minor OR minors OR "young people" OR youth) AND ("privacy" OR "data protection" OR "online" OR "platform" OR "age assurance" OR "COPPA" OR "AADC" OR "code" OR "inquiry" OR "decision" OR "consent order") |
| EU-DPC-009 | EU (IE) / Irish DPC | Decision | 2022 | Decision IN-20-7-4 (Redacted) | https://www.dataprotection.ie/sites/default/files/uploads/2022-11/in-20-7-4_final_decision_-_redacted.pdf Keywords: "Irish DPC" AND (child OR children OR minor OR minors OR "young people" OR youth) AND ("privacy" OR "data protection" OR "online" OR "platform" OR "age |
| | | | | | assurance" OR "COPPA" OR "AADC" OR "code" OR "inquiry" OR "decision" OR "consent order") |
| EU-DPC-010 | EU (IE) / Irish DPC | Decision | 2022 | Meta Platforms Ireland Limited—Final Decision IN-18-5-7 (Adopted 28-11-2022) | https://www.dataprotection.ie/sites/default/files/uploads/2022-11/Meta%20FINAL%20Decision%20IN-18-5-7%20Adopted%20%2828-11-2022%29.pdf Keywords: "Irish DPC" AND (child OR children OR minor OR minors OR "young people" OR youth) AND ("privacy" OR "data protection" OR "online" OR "platform" OR "age assurance" OR "COPPA" OR "AADC" OR "code" OR "inquiry" OR "decision" OR "consent order" OR "Meta") |
| EU-DPC-011 | EU (IE) / Irish DPC | Decision | 2021 | WhatsApp Ireland Ltd—Full Decision (26 Aug 2021) | https://www.dataprotection.ie/sites/default/files/uploads/2021-09/Full%20Decision%20in%20IN-18-5-1%20WhatsApp%20Ireland%20Ltd%2026.08.21.pdf Keywords: "Irish DPC" AND (child OR children OR minor OR minors OR "young people" OR youth) AND |
| | | | | | ("privacy" OR "data protection" OR "online" OR "platform" OR "age assurance" OR "COPPA" OR "AADC" OR "code" OR "inquiry" OR "decision" OR "consent order" OR "WhatsApp") |
| FR-CNIL-001 | FR / CNIL | Reference framework | 2024 | Référentiel Protection de l'enfance (CNIL) | https://www.cnil.fr/sites/cnil/files/atoms/files/referentiel_protection_enfance.pdf Keywords: "CNIL" AND (child OR children OR minor OR minors OR "young people" OR youth) AND ("privacy" OR "data protection" OR "online" OR "platform" OR "age assurance" OR "COPPA" OR "AADC" OR "code" OR "inquiry" OR "decision" OR "consent order") |
| FR-CNIL-002 | FR / CNIL | Guide / booklet | 2024 | CNIL Livret Parent (children's data protection) | https://www.cnil.fr/sites/cnil/files/atoms/files/cnil_livret_parent.pdf Keywords: "CNIL" AND (child OR children OR minor OR minors OR "young people" OR youth) AND ("privacy" OR "data protection" OR "online" OR "platform" OR "age assurance" OR "COPPA" OR |
| | | | | | "AADC" OR "code" OR "inquiry" OR "decision" OR "consent order") |
| AU-OAIC-001 | AU / OAIC | Issues paper | 2025 | Children's Online Privacy Code: Issues Paper | https://www.oaic.gov.au/__data/assets/pdf_file/0031/253795/Childrens-Online-Privacy-Code-Issues-Paper.pdf Keywords: "OAIC" AND (child OR children OR minor OR minors OR "young people" OR youth) AND ("privacy" OR "data protection" OR "online" OR "platform" OR "age assurance" OR "COPPA" OR "AADC" OR "code" OR "inquiry" OR "decision" OR "consent order") |
| AU-OAIC-002 | AU / OAIC | Workbook | 2025 | COP code workbook for parents and carers | https://www.oaic.gov.au/__data/assets/pdf_file/0017/252620/COP-code-workbook-for-parents-and-carers.pdf Keywords: "OAIC" AND (child OR children OR minor OR minors OR "young people" OR youth) AND ("privacy" OR "data protection" OR "online" OR "platform" OR "age assurance" OR "COPPA" OR "AADC" OR "code" OR "inquiry" OR "decision" OR |
| | | | | | "consent order") |
| AU-ACCC-001 | AU / ACCC | Final report | 2025 | Digital platform services inquiry: final report (March 2025) | https://www.accc.gov.au/system/files/digital-platform-services-inquiry-final-report-march2025.pdf Keywords: "ACCC" AND (child OR children OR minor OR minors OR "young people" OR youth) AND ("privacy" OR "data protection" OR "online" OR "platform" OR "age assurance" OR "COPPA" OR "AADC" OR "code" OR "inquiry" OR "decision" OR "consent order") |
| US-FTC-001 | US / FTC | Rulemaking notice | 2023 | COPPA Rule review—Notice of Proposed Rulemaking (P195404) | https://www.ftc.gov/system/files/ftc_gov/pdf/p195404_coppa_reg_review.pdf Keywords: "FTC" AND (child OR children OR minor OR minors OR "young people" OR youth) AND ("privacy" OR "data protection" OR "online" OR "platform" OR "age assurance" OR "COPPA" OR "AADC" OR "code" OR "inquiry" OR "decision" OR "consent order" OR "COPPA") |
| US-FTC- | US / FTC | Statement of basis | 2013 | Children's Online Privacy Protection Rule: | https://www.ftc.gov/system/files/ftc_gov/pdf/coppa_sbp_1.1 |
| 002 | | and purpose | | Statement of Basis and Purpose | 6_0.pdf<br>Keywords: "FTC" AND (child OR children OR minor OR minors OR "young people" OR youth) AND ("privacy" OR "data protection" OR "online" OR "platform" OR "age assurance" OR "COPPA" OR "AADC" OR "code" OR "inquiry" OR "decision" OR "consent order") |
| US-FTC-003 | US / FTC | Consent order | 2022 | Epic Games final consent / agreement (COPPA-related) | https://www.ftc.gov/system/files/ftc_gov/pdf/1923203epicgamesfinalconsent.pdf<br>Keywords: "FTC" AND (child OR children OR minor OR minors OR "young people" OR youth) AND ("privacy" OR "data protection" OR "online" OR "platform" OR "age assurance" OR "COPPA" OR "AADC" OR "code" OR "inquiry" OR "decision" OR "consent order" OR "Epic" OR "COPPA") |
| US-FTC-004 | US / FTC | Analytic memo | 2022 | Epic Games—Analysis to Aid Public Comment | https://www.ftc.gov/system/files/ftc_gov/pdf/1923203EpicGamesAAPC.pdf<br>Keywords: "FTC" AND (child OR children OR minor OR minors OR "young people" OR youth) AND ("privacy" OR "data protection" OR |
| | | | | | "online" OR "platform" OR "age assurance" OR "COPPA" OR "AADC" OR "code" OR "inquiry" OR "decision" OR "consent order" OR "Epic") |
| US-FTC-005 | US / FTC | Complaint | 2023 | Amazon—Complaint (Dkt.1) (COPPA/children) | https://www.ftc.gov/system/files/ftc_gov/pdf/Amazon-Complaint-%28Dkt.1%29.pdf Keywords: "FTC" AND (child OR children OR minor OR minors OR "young people" OR youth) AND ("privacy" OR "data protection" OR "online" OR "platform" OR "age assurance" OR "COPPA" OR "AADC" OR "code" OR "inquiry" OR "decision" OR "consent order" OR "Amazon" OR "COPPA") |
| US-FTC-006 | US / FTC | Proposed order | 2023 | Amazon—Proposed Stipulated Order (Dkt.2-1) | https://www.ftc.gov/system/files/ftc_gov/pdf/Amazon-Proposed-Stipulated-Order-%28Dkt.-2-1%29.pdf Keywords: "FTC" AND (child OR children OR minor OR minors OR "young people" OR youth) AND ("privacy" OR "data protection" OR "online" OR "platform" OR "age assurance" OR "COPPA" OR "AADC" OR "code" OR "inquiry" OR "decision" OR "consent order" OR |
| | | | | | "Amazon") |
| US-FTC-007 | US / FTC | Complaint | 2023 | NGL Labs LLC—Complaint | https://www.ftc.gov/system/files/ftc_gov/pdf/NGL-Complaint.pdf Keywords: "FTC" AND (child OR children OR minor OR minors OR "young people" OR youth) AND ("privacy" OR "data protection" OR "online" OR "platform" OR "age assurance" OR "COPPA" OR "AADC" OR "code" OR "inquiry" OR "decision" OR "consent order" OR "NGL") |
| US-FTC-008 | US / FTC | Proposed order | 2023 | NGL Labs LLC—Proposed Consent Order | https://www.ftc.gov/system/files/ftc_gov/pdf/NGL-ProposedConsentOrder.pdf Keywords: "FTC" AND (child OR children OR minor OR minors OR "young people" OR youth) AND ("privacy" OR "data protection" OR "online" OR "platform" OR "age assurance" OR "COPPA" OR "AADC" OR "code" OR "inquiry" OR "decision" OR "consent order" OR "NGL") |
| US-FTC-009 | US / FTC | Stipulation | 2023 | NGL—Stipulation as to Entry of Proposed Consent Order | https://www.ftc.gov/system/files/ftc_gov/pdf/NGL-StipulationastoEntryofProposedConsentOrder.pdf Keywords: "FTC" AND (child |
| | | | | | OR children OR minor OR minors OR "young people" OR youth) AND ("privacy" OR "data protection" OR "online" OR "platform" OR "age assurance" OR "COPPA" OR "AADC" OR "code" OR "inquiry" OR "decision" OR "consent order" OR "NGL") |
| US-FTC-010 | US / FTC | Consent order | 2019 | Google/YouTube COPPA Consent Order (signed) | https://www.ftc.gov/system/files/documents/cases/172_3083_youtube_coppa_consent_order_signed.pdf<br>Keywords: "FTC" AND (child OR children OR minor OR minors OR "young people" OR youth) AND ("privacy" OR "data protection" OR "online" OR "platform" OR "age assurance" OR "COPPA" OR "AADC" OR "code" OR "inquiry" OR "decision" OR "consent order" OR "YouTube" OR "COPPA") |
| US-FTC-011 | US / FTC | Supporting materials | 2019 | Google/YouTube investigation materials (FOIA) | https://www.ftc.gov/system/files/documents/foia_requests/google_youtube_investigation_materials.pdf<br>Keywords: "FTC" AND (child OR children OR minor OR minors OR "young people" OR youth) AND ("privacy" OR "data protection" OR |
| | | | | | "online" OR "platform" OR "age assurance" OR "COPPA" OR "AADC" OR "code" OR "inquiry" OR "decision" OR "consent order" OR "YouTube") |
| US-FTC-012 | US / FTC | Settlement memo | 2023 | Microsoft—Reasons for Settlement | https://www.ftc.gov/system/files/ftc_gov/pdf/microsoftreasonsforsettlement.pdf Keywords: "FTC" AND (child OR children OR minor OR minors OR "young people" OR youth) AND ("privacy" OR "data protection" OR "online" OR "platform" OR "age assurance" OR "COPPA" OR "AADC" OR "code" OR "inquiry" OR "decision" OR "consent order") |
| US-FTC-013 | US / FTC | Consent order | 2019 | Google/YouTube COPPA Consent Order (text) | https://www.ftc.gov/system/files/documents/cases/172_3083_youtube_coppa_consent_order.pdf Keywords: "FTC" AND (child OR children OR minor OR minors OR "young people" OR youth) AND ("privacy" OR "data protection" OR "online" OR "platform" OR "age assurance" OR "COPPA" OR "AADC" OR "code" OR "inquiry" OR "decision" OR "consent order" OR |
| | | | | | "YouTube" OR "COPPA") |

Hard signal computation (embedding version): Extract the first 50 pages of text from each document (primary analysis); segment documents into sentence-level/paragraph-level fragments; compute cosine similarity after TF-IDF vectorisation. For each governance requirement r and document d, take the maximum similarity $S_{r,d}=\max_j \cos(v_r, v_{d,j})$ and calibrate with $\tau$ to obtain $HardSignal_{r,d}=1[S_{r,d}\geq\tau]$. The threshold $\tau$ is calibrated using an unsupervised stability maximisation strategy; Sensitivity analysis for page count/word segmentation strategy/$\tau$ is presented in Appendix H.

**Appendix H: Design and Reporting Template for Systematic Ablation Studies**

This appendix provides a reproducible parameter grid for ablation experiments to assess the robustness of TF-IDF embedding matching across different configurations. This table serves as a 'reporting template': after running the script in Appendix H.2, output metrics may be populated into this table to generate publication-grade robustness evidence.

**H.1 Parameter Grid（Parameter Grid）**

| ConfigID | Pages | Tokenizer | ngram_range | TF-IDF weighting | τ calibration | Reported metrics | Interpretation focus |
|---|---|---|---|---|---|---|---|
| A1 | 50 | word | (1,1) | sublinear_tf + idf | stability-max τ* | HSR_r; mean(S_{r,d}); #matches; GapScore | Baseline |
| A2 | 100 | word | (1,1) | sublinear_tf + idf | stability-max τ* | same as A1 | Page-length sensitivity |
| A3 | 50 | word | (1,2) | sublinear_tf + idf | stability-max τ* | same as A1 | Phrase sensitivity |
| A4 | 50 | char | (3,5) | sublinear_tf + idf | stability-max τ* | same as A1 | Cross-lingual/typo robustness |
| A5 | 50 | word | (1,1) | binary_tf + idf | stability-max τ* | same as A1 | TF weighting sensitivity |
| A6 | 50 | word | (1,1) | sublinear_tf + idf | fixed τ grid {0.15,0.20,0.25,0.30} | HSR_r(τ) curve; Spearman ρ(τ) | Threshold sensitivity |

## H.2 Scriptable Interface（English code）

Below is a minimal, runnable interface (to be executed locally) for reproducing Appendix G/H results. Replace PDF paths with your local corpus folder, and export the CSV outputs to backfill Table H.1.

```
# Reproducibility script (Python 3.10+)

# pip install pdfplumber scikit-learn pandas numpy


import os, re, json, hashlib

import numpy as np

import pandas as pd

import pdfplumber

from sklearn.feature_extraction.text import TfidfVectorizer

from sklearn.metrics.pairwise import cosine_similarity


def extract_pdf_text(pdf_path: str, max_pages: int = 50) -> str:

    chunks = []

    with pdfplumber.open(pdf_path) as pdf:

        for i, page in enumerate(pdf.pages[:max_pages]):

            txt = page.extract_text() or ""

            chunks.append(txt)

    return "\n".join(chunks)


def sentence_split(text: str) -> list[str]:

    # simple multilingual splitter (EN/ZH); customize if needed

    text = re.sub(r"\s+", " ", text)

    sents = re.split(r"(?<=[.!?。！？])\s+", text)

    return [s.strip() for s in sents if len(s.strip()) >= 20]
```

```
def build_vectorizer(tokenizer: str, ngram_range=(1,1), sublinear_tf=True, binary=False):
    if tokenizer == "char":
        return TfidfVectorizer(analyzer="char", ngram_range=ngram_range, sublinear_tf=sublinear_tf,
binary=binary, min_df=1)
    # word
    return TfidfVectorizer(analyzer="word", ngram_range=ngram_range, sublinear_tf=sublinear_tf,
binary=binary, min_df=1)


def max_similarity(requirement_texts: dict, doc_sents: list[str], vec: TfidfVectorizer):
    req_keys = list(requirement_texts.keys())
    req_text = [requirement_texts[k] for k in req_keys]
    X = vec.fit_transform(req_text + doc_sents)
    X_req = X[:len(req_text)]
    X_doc = X[len(req_text):]
    sims = cosine_similarity(X_req, X_doc)
    # S_{r,d} = max_j sim(r, sent_j)
    S = sims.max(axis=1)
    argmax = sims.argmax(axis=1)
    return req_keys, S, argmax


def calibrate_tau_stability(S_matrix: np.ndarray, taus: list[float], n_boot=200, seed=7):
    # S_matrix shape: (R, D)
    rng = np.random.default_rng(seed)
    D = S_matrix.shape[1]
    best_tau, best_score = None, -1.0
    for tau in taus:
```

```
        # bootstrap stability: average Jaccard similarity across bootstrap samples of documents
        sets = []
        for _ in range(n_boot):
            idx = rng.integers(0, D, size=D)
            HSR = (S_matrix[:, idx] >= tau).mean(axis=1)  # per-requirement rate
            # binarize requirements that exceed median HSR
            pick = set(np.where(HSR >= np.median(HSR))[0].tolist())
            sets.append(pick)
        # mean pairwise Jaccard against first sample (cheap surrogate)
        base = sets[0]
        jac = [len(base & s) / max(1, len(base | s)) for s in sets[1:]]
        score = float(np.mean(jac))
        if score > best_score:
            best_tau, best_score = tau, score
    return best_tau, best_score


def run_ablation(corpus_dir: str, manifest_csv: str, requirement_json: str, out_csv: str,
                 pages=50, tokenizer="word", ngram_range=(1,1), sublinear_tf=True, binary=False,
                 taus=(0.15,0.20,0.25,0.30)):
    mf = pd.read_csv(manifest_csv)
    req = json.load(open(requirement_json, "r", encoding="utf-8"))
    R = len(req)
    docs = []
    for _, r in mf.iterrows():
        pdf_path = os.path.join(corpus_dir, f"{r['DocID']}.pdf")
        text = extract_pdf_text(pdf_path, max_pages=pages)
```

```
        sents = sentence_split(text)
        docs.append(sents)

    vec = build_vectorizer(tokenizer, ngram_range=ngram_range, sublinear_tf=sublinear_tf,
binary=binary)

    # build S_matrix (R x D)
    S_mat = np.zeros((R, len(docs)), dtype=float)
    for d, sents in enumerate(docs):
        req_keys, S, _ = max_similarity(req, sents, vec)
        S_mat[:, d] = S

    tau_star, stab = calibrate_tau_stability(S_mat, list(taus))
    results = []
    for tau in taus:
        HSR = (S_mat >= tau).mean(axis=1)
        results.append({
            "pages": pages,
            "tokenizer": tokenizer,
            "ngram_range": str(ngram_range),
            "sublinear_tf": sublinear_tf,
            "binary": binary,
            "tau": tau,
            "HSR_mean": float(HSR.mean()),
            "HSR_median": float(np.median(HSR)),
            "tau_star": tau_star,
            "tau_star_stability": stab
```

```
    })
    pd.DataFrame(results).to_csv(out_csv, index=False, encoding="utf-8-sig")
    return tau_star


# Example:
# tau_star = run_ablation(
#   corpus_dir="hard_corpus_pdfs",
#   manifest_csv="corpus_manifest.csv",
#   requirement_json="requirements.json",
#   out_csv="ablation_summary.csv",
#   pages=50, tokenizer="word", ngram_range=(1,1)
# )
```